\documentclass{steme}

\usepackage{txfonts}

\def\typeofarticle{to appear STEM Education Journal (2025) \\
Research article}

\makeatletter
\renewcommand{\@biblabel}[1]{#1\hfill \hspace{-0.2cm}}
\makeatother

\usepackage{cite}
\begin{document}

\title{Hands-on STEM learning experiences using digital technologies}

\author{Gaia Fior, Carlo Fonda and Enrique Canessa\corrauth}


\address{{SciFabLab, International Centre for Theoretical Physics (ICTP), 34151 Trieste, Italy}}

\corraddr{canessae@ictp.it.}

\begin{abstract}
The facilitation of STEM education can be enhanced by the provision of opportunities
for learners through the utilization of tangible and visual examples that lead to gain
a better understanding of STEM topics. The objective of this work was to present
an account of our experiences and activities with this
novel approach carried out on schools in the Italian FVG
region and also shown and tested on large, free public activities like Maker Faires 
and Science Picnics. The projects and experiences discussed---in which students
develop a range of core competencies such as creativity, critical thinking,
experimentation, prototyping, collaboration and problem-solving---include tangible 
complex 3D printed structures, large microcontroller board replicas
and the visualization of wind dynamics and tiny invisible elementary particles,
among others. These hands-on experiences demonstrate the benefits on the use of
digital fabrication technologies implemented within a FabLab for STEM learning.
We have identified and proposed a set of valid examples for possible
engagements that are beyond today’s standard education and may provide more
authentic learning to nurture 21st-century skills.
\end{abstract}


\maketitle

\section{Introduction}
\label{sec:1}

The present approach to develop hands-on science, technology, engineering, and
mathematics (STEM) learning experiences using digital technologies is
motivated by the 2030 UN agenda for sustainable development, which provides
a global framework for progress and sustainable development. In order to achieve
these goals, it is essential to implement innovative strategies that will lead to
improvements in STEM education and the creation of opportunities for more
scholars. In particular, the Sustainable Development Goal 4 of the UN 2030
Agenda aims to \emph{``ensure inclusive and equitable quality education and promote
lifelong learning opportunities for all"} \cite{bib1}.

In the last decade, online education has transformed society to new levels in the
way people teach and learn \cite{bib2}. Additionally, a number of other significant 
digital innovations are currently in place which collectively aim to facilitate the
achievement of the 2030 UN goals. Examples are the use of 3D printing for the
production of Mathematics and Physics educational structures, the application of
microcontrollers and tiny computers for sensor-based experiments \cite{bib3}. These 
are becoming increasingly popular, which are helping to stimulate scholars' curiosity
and encourage a more profound understanding of specific subjects. Thus, today, learning 
is facilitated under a personal context in which scholars 
engage in reflection, imagination, comprehension and comparison of solutions through 
the use of tangible examples. The implementation of digitally controlled 
technologies for tangible and visual STEM learning is enabling scholars to access 
the highest quality education.

The skills needed to make good on the UN’s agenda need to be nurtured over many years, 
and to start working in (elementary, primary and secondary) schools is certainly one of 
the necessary steps. It is also necessary to identify and test 
a set of valid examples for possible engagements that are beyond today’s 
standard education, which usually include videos and other multimedia forms, 
and provide an authentic learning to nurture 21st-century skills.
Digital Fabrication Laboratories (FabLabs) can help in the process, since they
are a global network of local labs, enabling invention by providing access to tools 
for rapid digital fabrication. In these places,
the transformation of an idea into a tangible object is possible, making it a
potential pedagogical tool in STEM education. The establishment of FabLabs in
academia can play a relevant role as a resource for the creation of pedagogical
objects and devices to support sustainable development within the educational
context.

Implementing STEM learning in classrooms is a challenge, and therefore
it is important to pay attention to problems in current STEM learning, and see 
how hands-on experiences or the use of digital technologies can solve such problems.
Some important issues in current STEM education include, for example, finding the necessary 
financial funding to obtain resources for its implementation as well as how
the load on the educators' work can be optimized. Both of these problems can
benefit by the use of digital fabrication technologies implemented within a FabLab 
for STEM learning. There are a myriad of digital resources and tools that 
can be downloaded for free on the internet. Instructors should become a guide in this process 
and improve learning efficiency. However, teachers frequently lack digital expertise 
and receive little training in this context.  

Engaging educational objects and materials to inspire students
with demos and do-it-yourself (DIY) constructions to better grasp complex STEM concepts
can be developed 
in collaboration with a local FabLab community with ongoing technical training and advice.
Collaborations between schools and FabLab partners can lead to unique learning 
opportunities, such 
as developing the needed objects and examples. A visit to these open space facilities or the co-organization 
of informative workshops hosted at FabLabs open the path to solve some of the problems of implementation
in current STEM learning. In the end, all these efforts will be rewarded giving 
scholars the possibility to develop essential 21st-century skills. 
Integrating STEM education into schools can then be a reality. 

Teachers need to find available FabLabs and Makerspaces in their areas and be aware of the 
possibilities offered within these offices. Being updated with new technical progress, 
they help to open new opportunities and connections to integrate STEM subjects within the school curriculum.
A better perception of STEM subjects by teachers and students through hands-on activities, 
applications and examples will help to gain interest in STEM subjects and can become suitable 
for all kinds of students---without any gender distinction and concrete opportunities for inclusion. 
By exposing students to visual and tangible STEM education, they gain extra valuable insights into 
problem-solving and teamwork. STEM lessons should cater to diverse student needs by visual and 
tangible education designed in FabLabs.

In this work, we describe some of our experiences applying a hands-on and visual
approach to STEM education, developed in our Scientific Fabrication Laboratory
(SciFabLab) which is equipped with the necessary facilities to build specific devices
for that purpose. The specifications of these devices are openly shared online.
We argue that placing learning literally in the hands of the younger generations is
a more effective approach for all, especially when it does not pose any special technical
requirements on the teachers' side.

Educational efforts toward a better understanding of science through the
utilization of tangible and visual examples, as those illustrated and
explained in our paper, are only being explored recently because of the
new available digital technologies.
The selection of projects and hands-on experiences discussed in our paper
aims to demonstrate and promote the benefits of the use of digital fabrication
technologies implemented within a FabLab for STEM learning.
Ours outcomes are based on our own experiences for science dissemination
and STEM education, which were not only carried out in some schools in
Italy but also at large, free public events like Maker Faires and Science
Picnics.

Furthermore, our objective is also to develop techniques and processes that will
enable new generations of students to participate fully in the educational process,
in a manner that can be inclusive to incorporate individuals with physical 
disabilities \cite{bib4}. As discussed in this work, STEM activities can be adapted
to accommodate specific needs with the objective of overcoming some students'
physical limitations and impairments. 

\section{Previous works}
\label{sec:2}

Some 21 st-century skills are difficult to teach only by traditional methods,
i.e., without the use of highly sophisticated information and communications
technologies (ICT). However, further studies on the design, creation and
implementation of new instruments in which the teaching-learning process can be
more personalized is still needed, see a review in \cite{bib5}. The use of ICT
has been recurrent in the last decade, especially due
to the massive diffusion of portable PCs, tablets and smartphones \cite{bib2}.
Nevertheless, alternative digital technologies available at FabLabs can be
employed to assist in the development of innovative pedagogical approaches
that are accessible to all students and that facilitate their understanding on 
the relevance of scientific concepts \cite{bib3}.

FabLabs facilitate the implementation of tangible and visual STEM
learning, thereby opening up new possibilities. The following sections will present
a selection of our projects and experiences, in which students develop a range of
core competencies, including creativity, critical thinking, experimentation and
collaboration, prototyping and problem-solving. A discussion
to better understand the influence of Makerspaces and FabLabs on today's
innovation processes can be found in \cite{bib6}.
STEM education opportunities for everyone via the implementation of
Makerspaces and FabLabs in vulnerable and marginal communities is discussed
in \cite{bib7}. On the other hand, the potential aspects of FabLabs 
in relation to traditional books as a pedagogical tool, and how teachers perceive these
tools for teaching Chemistry to their students, was investigated in \cite{bib8}.

The evidence gathered from these implementations lend support to STEM learning 
practices via FabLabs. We have verified that these new visual and tangible practices 
become a rewarding and meaningful experience for all participants.
These novel implementations can also help to enlarge adoption of 
educational digital tools in local communities and surrounding areas 
through word of mouth among students and teachers.

\section{Some STEM prototypes for all}
\label{sec:3}

The following section presents a selection of tangible and visual examples of
STEM devices that can be reproduced at any FabLab. FabLabs are typically
equipped with low-cost, state-of-the-art, and versatile computer-controlled rapid
prototyping tools including 3D printers and scanners, laser engraving, and CNC
cutting machines driven by open-source software. FabLabs provide a working
space for invention, creativity, and resourcefulness, and they are strongly rooted in
a local community of makers. STEM students can learn more effectively through
practice and be more productive through experience with the prototypes
produced in these FabLabs.

The aim of our FabLab is also to disseminate scientific knowledge and to provide
opportunities to all class of students to engage with STEM. The outcomes
reported here are the output of experiences derived from research activities
conducted at the ICTP Scientific Fabrication Laboratory of Trieste, Italy, over the
past few years. These experiences were conducted as hands-on, one-off learning 
experiences in some schools in Italy and in activities during events such as 
Science Picnics and Maker Faires.

\subsection{3D STEM objects}
\label{subsec:3a}

The current state of low-cost 3D printing technologies is such
that they have reached a level of maturity that allows for their unlimited potential
to be realized in the field of STEM education. The capacity to reproduce 3D
objects of varying complexity, ranging from intricate mathematical structures up
to sophisticated biological specimens, has not only benefited schools and higher-education institutions in countries with limited scientific infrastructure to sustain
their STEM education programs, but it has also facilitated the rapid growth of digital
STEM content around the world.

Typically, 3D printing with FDM (fused deposition modeling) technology is
conducted using spools of thermo-plastic filament like polylactic acid (PLA)---an
environmentally friendly material derived from corn starch. A number of free or
proprietary tools for 3D modeling and printing are available on the internet, facilitating
the work required to design and print the objects.

\begin{figure}[ht]
    \centering
        \includegraphics[width=0.367\textwidth]{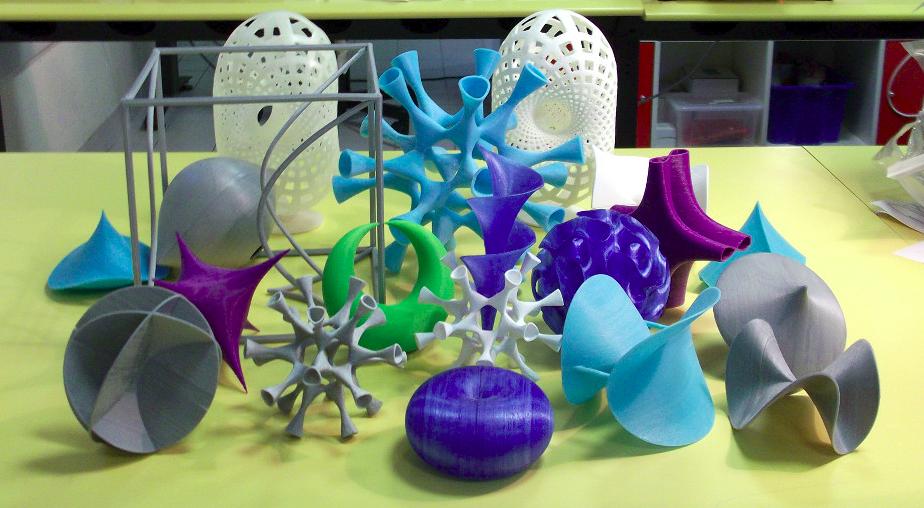} 
        \includegraphics[width=0.2\textwidth]{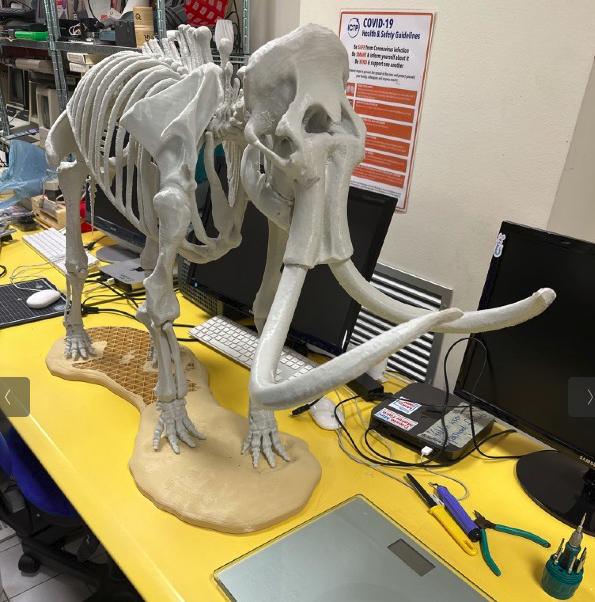} 
\caption{Left: Examples of complex 3D printed mathematical structures. Right: Scale model
of a giant, extinct elephantid genus Mammuthus.}
\label{fig:1}       
\end{figure}

\subsection{Highly visual cloud chamber---\emph {Physics in action!}}
\label{subsec:3d}

A facility for studying tiny, invisible elementary particles that enter our atmosphere
from the outer space (cosmic rays) or that are generated by the decay of radioactive
elements within the earth itself, and then pass right through our bodies, is the
well-known apparatus known as Wilson’s cloud chamber. This apparatus was
invented in the early 1900s by Charles Wilson as one of the first particle
detectors. The majority of these particles are generated by the decay of natural
radon found in the air (itself a product of radioactive elements decay in the
ground) while a certain percentage, which varies with altitude, does originate from
the cosmic radiation we receive from outer space (primary and secondary
cosmic rays).

\begin{figure}[ht]
    \centering
        \includegraphics[width=0.224\textwidth]{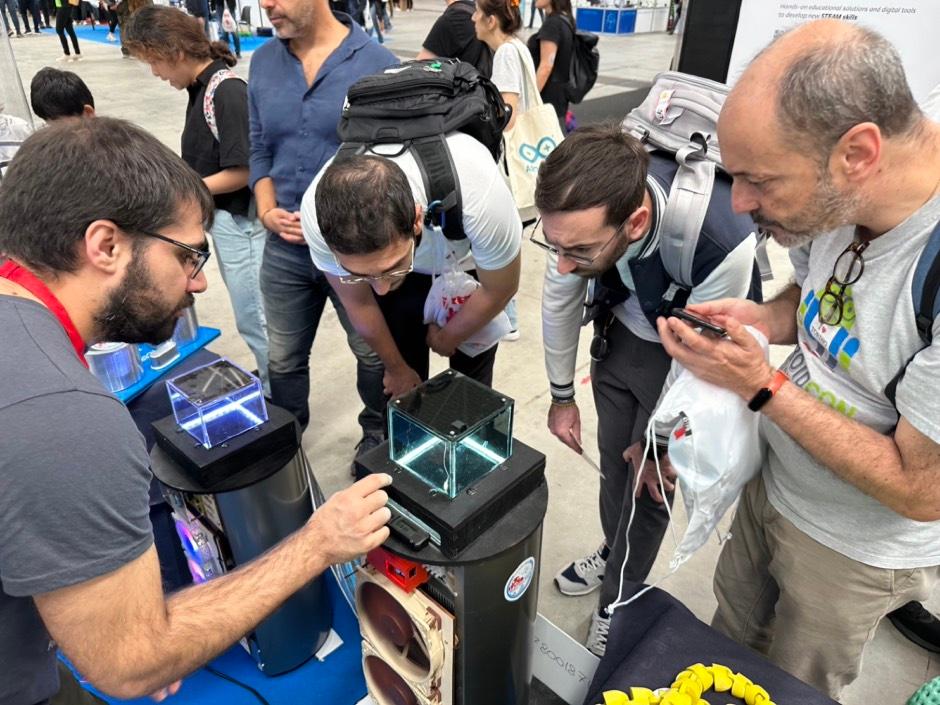}
        \includegraphics[width=0.3\textwidth]{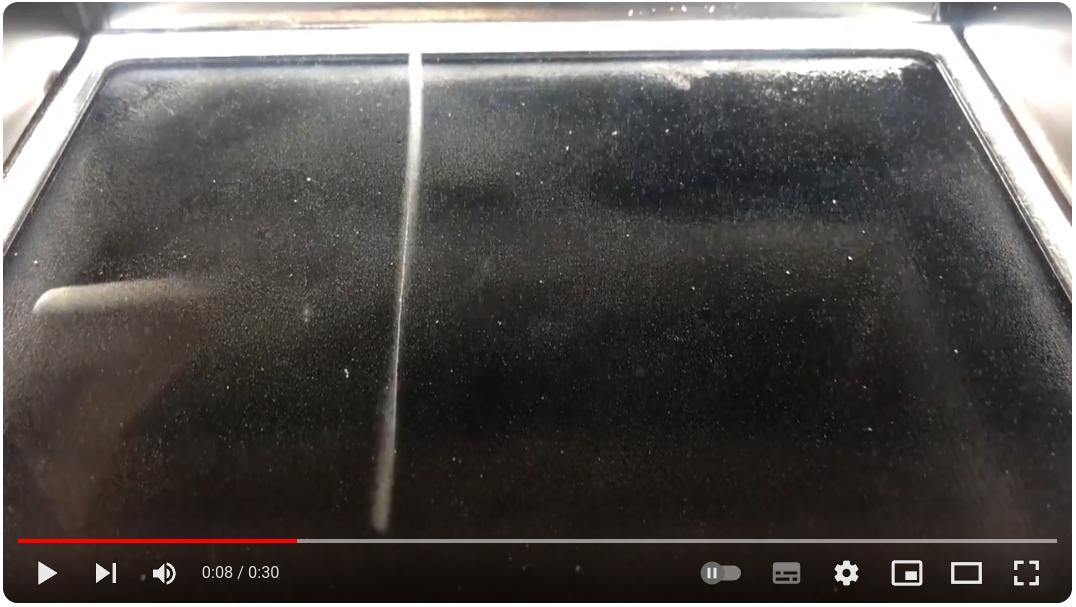}  
\caption{Typical particle tracks in a cloud chamber device (left). the full video is
available at https://www.youtube.com/watch?v=IwyLzqrRJbw. Right: Public interest 
on the DIY cloud chamber.}
\label{fig:5}       
\end{figure}

To build a cloud chamber for educational purposes, one needs a transparent box
with a black bottom metal tray maintained at a temperature of approximately
-30 degrees Celsius, accompanied by the presence of a stable cloud of
minute droplets of isopropyl alcohol, typically generated by condensation from a
liquid state. After a period of time, the top lid of the chamber, which has been
soaked in warm alcohol, produces a cloud that reaches the cold metal bottom.
When a charged particle passes through at a high speed, it generates a visible
track of larger droplets by a mechanism known as coalescence. The droplets of
alcohol can then be detected due to their ability to diffuse the light emitted 
by the strips of LEDs positioned on the sides, as illustrated in Figure \ref{fig:5}.

The particles that can be observed in a cloud chamber can be broadly classified
into three categories: alpha particles, beta particles, and muons. Alpha particles
are composed of two protons and two neutrons. They have a short, fat track and
are produced by the decay of radon. Beta particles are electrons (or anti-electrons, 
a.k.a. positrons), usually following long, straight paths or zigzag paths,
depending on the energy and momentum of the particles. Muons are particles
emitted from the decay of secondary cosmic rays and can be observed as short,
vertical tracks. Further details on the construction and operation of a cloud
chamber can be found in \cite{bib9}.

\subsection{Galileo and Leonardo mechanical clocks}
\label{subsec:3e}

\begin{figure}[ht]
    \centering
        \includegraphics[width=0.43\textwidth]{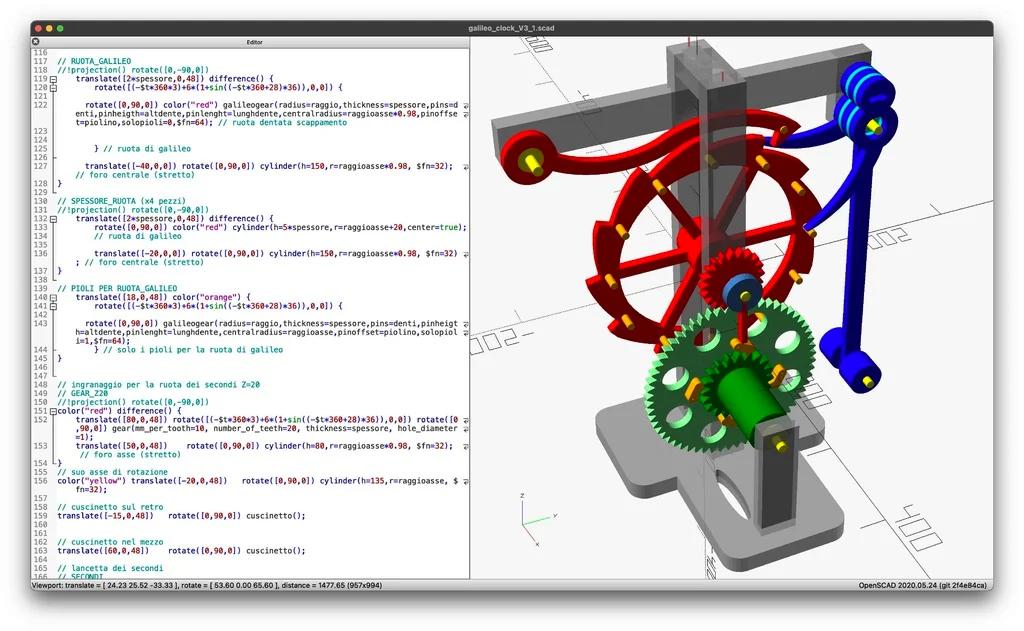}
        \includegraphics[width=0.15\textwidth]{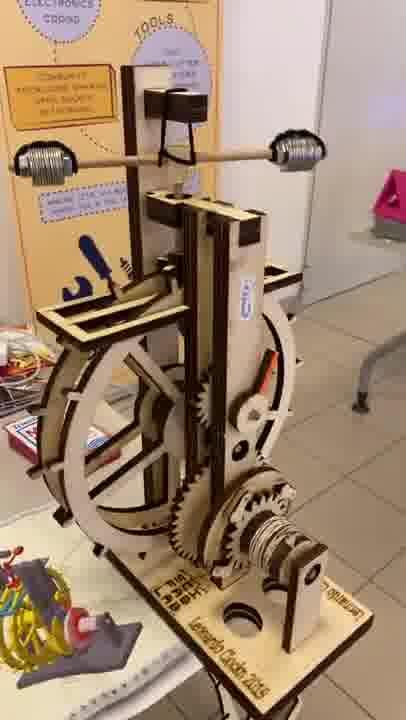}  
\caption{OpenSCAD design of a Galileo mechanical clock (left). The video demonstration is
available at: https://www.youtube.com/watch?v=a6B1bxgxAc4.
Right: Leonardo mechanical clock. The video demostration is available at:
https://www.youtube.com/shorts/g4hkIHu89kE.
}
\label{fig:6}       
\end{figure}

The first application of a pendulum in a clock using an escapement system is
attributed to Galileo Galilei from his original drawings. Following the discovery 
of the isochronism of a pendulum by Galileo's observation of a swinging lamp in
Pisa Cathedral (dating back to 1600), a working gravity-powered clock can be
made on a laser cutter with the addition of standard bearings and a few parts
made from plywood. As shown in Figure \ref{fig:6} (left), the design of a Galileo mechanical
clock can be written using OpenSCAD---a free, fully customized software.

Galileo’s clock mechanism uses gravitational potential energy as its power
source, and carries two arcs; a small, lower arc that receives the impulse, while an
upper, wide arc lifts a detent. This model has been designed in collaboration with
the “Amici dell’orologeria Pesarina” association, engaged for many years in the
dissemination of knowledge concerning watchmaking, a tradition present in the
area of Pesariis (an ancient village in the mountains of the Friuli Venezia Giulia (FVG) region,
Italy) since 1600, and its dissemination in local schools. The request was to produce 
a model escapement that could be made cheaply and then to introduce a kit into
secondary schools so that students could experience the mechanisms first-hand.
The files for this build can be downloaded freely (https://www.thingiverse.com/
thing:5549040) and can be further customized in OpenSCAD easily.

The Leonardo Da Vinci clock depicted in Figure \ref{fig:6} (right) is another wooden
pendulum clock that can be constructed in a FabLab for educational purposes
(https://www.thingiverse.com/thing:4328105). This clock model is powered by 
a horizontal rotary pendulum, and utilizes the weight contained in a canister and
some gears to measure time. The canister has stored potential energy gained by
raising it up against gravity. The potential energy is converted to kinetic energy
allowing the gears to move their parts. By adjusting the weight and balance of the
pendulum, the clock's speed can then be controlled.

\subsection{Giant Galton board}
\label{subsec:3f}

\begin{figure}[ht]
    \centering
        \includegraphics[width=0.17\textwidth]{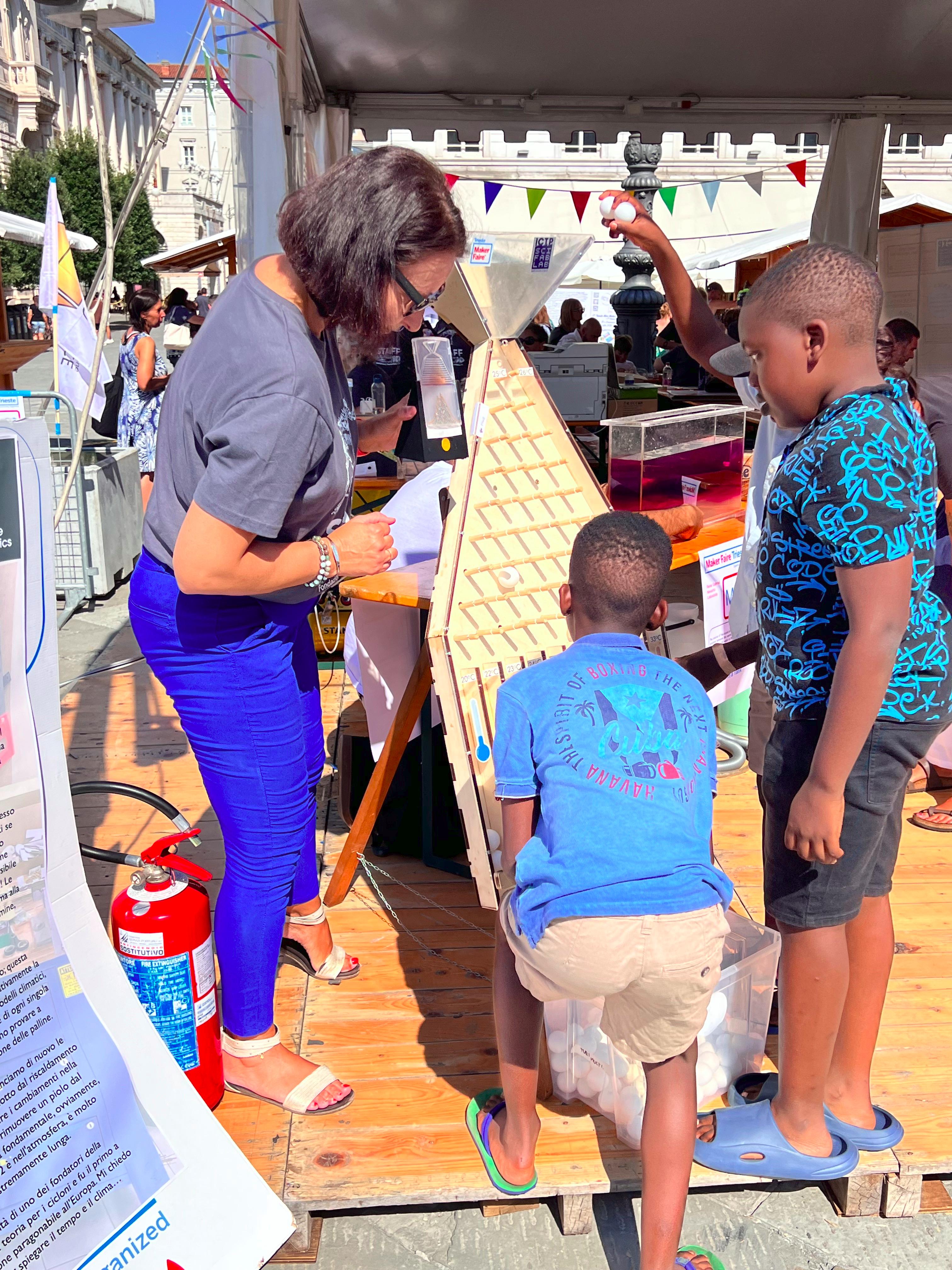}
        \includegraphics[width=0.341\textwidth]{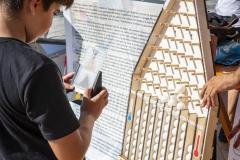}  
\caption{Giant Galton board of descent ping-pong balls.}
\label{fig:7}       
\end{figure}

A large Galton board allows students to visualize the preferred trajectories of
falling (ping-pong) balls while being fed into the device from the top at arbitrary
rates. Due to the differing initial conditions established by the participants, the
gravitational force and the presence of ordered, fixed obstacles along their paths, the
individual balls experience random elastic collisions as they descend. The falling
balls' positions will not concentrate at specific locations. Instead, they will be
characterized by a distribution (with one or more peaks) resulting from the
influence of the stochastic effects due to the presence, or absence, of the
obstacles. The spatial distribution of the falling balls with homogeneous obstacles
is observed to exhibit a symmetric bell-shaped (Gaussian) curve, see Figure~\ref{fig:7}.

The observation of individual ball trajectories represents the dynamics of a single
ensemble member, whereas the bell statistical distribution of the final balls'
positions represents the total ensemble average for the system. This device can
serve as a learning platform for explaining weather forecast models, as they use
the equations governing the atmosphere in conjunction with a stochastic
algorithm called “Monte Carlo” to generate a weather forecast (corresponding to
the resulting distribution of the balls). The model, as well as the device, exhibits a
high degree of sensitivity to its initial conditions (of the dropped balls), where 
even minor initial perturbations give rise to divergent weather predictions. The
presence or absence of the obstacles correspond to the various parameters of
the underlying model, and they can be moved to “tune” the results.

\subsection{Augmented Reality Sandbox}
\label{subsec:3g}

The Augmented Reality (AR) Sandbox is an educational hands-on exhibit
developed by the University of California, Davis, USA. It combines a box filled
with real sand with virtual topography and simulated water. The apparatus is
constructed using a Microsoft Kinect 3D camera, in addition to a video projector
and a Linux PC to achieve extensive real-time computer simulations and
visualization. The AR Sandbox allows the reconstruction of geographical models
through the shaping of the sand. These models are then augmented by an
elevation color map, topographic contour lines, and simulated liquid water.

\begin{figure}[ht]
    \centering
        \includegraphics[width=0.22\textwidth]{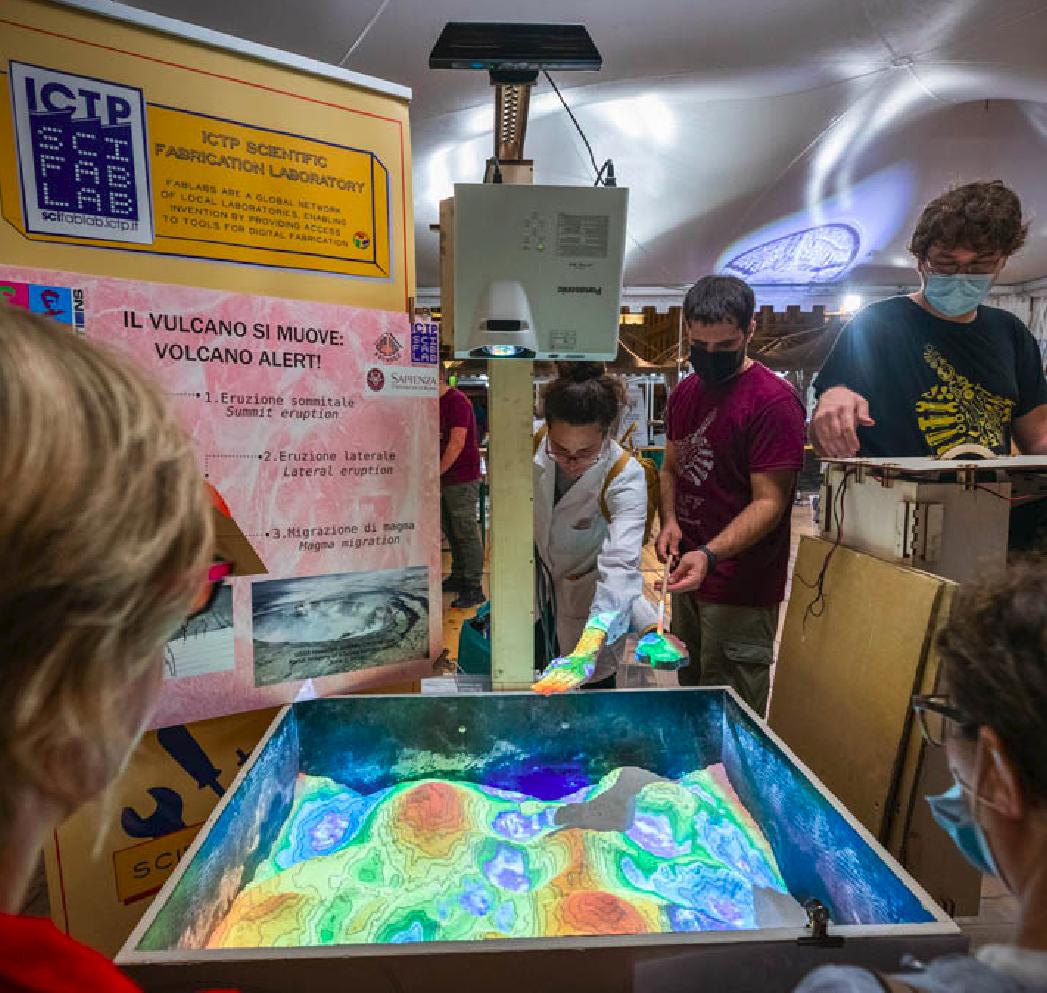}
        \includegraphics[width=0.292\textwidth]{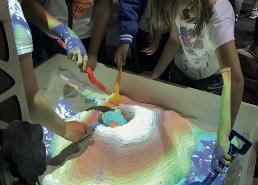}  
\caption{The Sandbox unit (left) which displays a mountain with a crater 
lake, surrounded by several smaller lakes (right). This tool is an invaluable
resource for educators and learners alike, facilitating the acquisition 
of knowledge and understanding in the domains of topography, geography,
natural sciences and water flow.}
\label{fig:8}       
\end{figure}

The device, illustrated in Figure \ref{fig:8}, has been employed for science communication
projects involving experiments on geological hazards and volcano alerts. The
system is useful for illustrating concepts related to modeling volcanoes, including
the processes involved in volcano deformation and the study of magma
reservoirs.

A variety of hands-on experiments can be conducted, such as placing a small,
inflated balloon beneath the sand. Deflating the balloon can then demonstrate the
effect of a collapsing volcano magmatic chamber. It is also possible to illustrate
the impact of the rising sea level on coastal areas: for instance, Easter Island,
located in the southeastern Pacific Ocean, is highly vulnerable to the effects 
of climate change, and this can be visually demonstrated with great clarity by
varying digitally the height of the sea level.

\subsection{2D-to-3D image conversion via Artificial Intelligence (AI) model}
\label{subsec:3h}

3D visualization has the potential to enhance the understanding of STEM among
young students, thereby facilitating more effective learning. Virtual learning 
opens up new worlds in STEM education, particularly in the context of abstract
concepts and the development of specific experiences \cite{bib10}. Virtual learning can
facilitate the exploration of the interior of a human cell, traverse planetary
surfaces, or navigate subterranean magma chambers. These are examples of the
``intangible world" that can be experienced in virtual learning environments. 
In order for this to be achieved, it is necessary to recreate a realistic 3D
environment. It would be even more beneficial if the viewer were able to perceive
this content more naturally, i.e., without the need for special 3D glasses.

\begin{figure}[ht]
    \centering
        \includegraphics[width=0.2\textwidth]{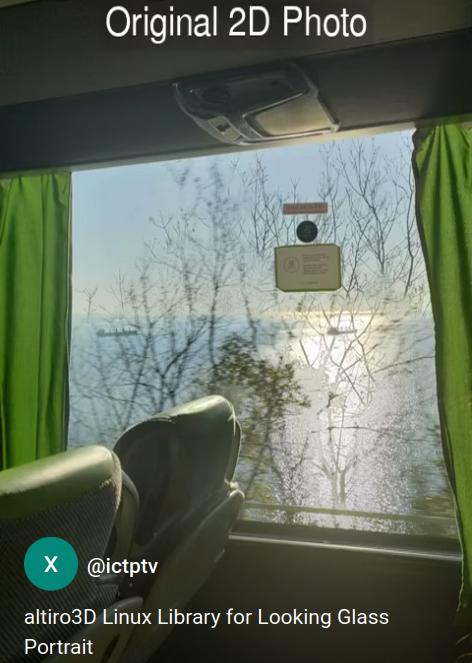}
        \includegraphics[width=0.2\textwidth]{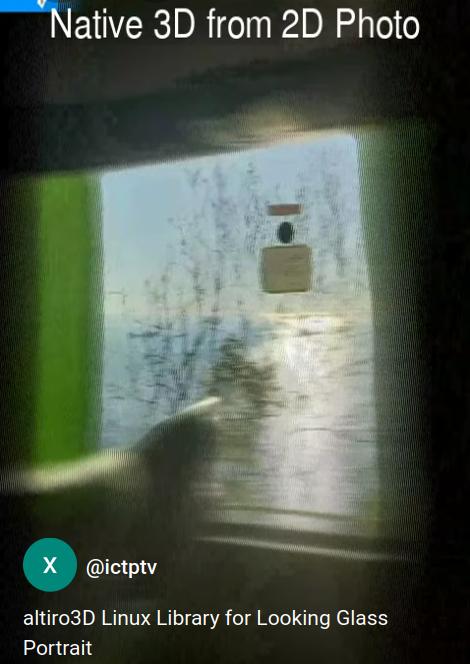}  
        \includegraphics[width=0.197\textwidth]{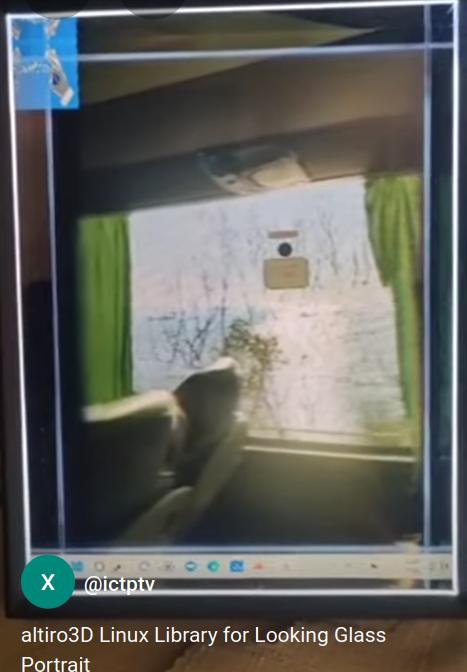} 
\caption{Outputs of the ``altiro3D" algorithm to represent reality.
Left: Original, single RGB image. Center: Multiview synthesis of a single image carried out 
through AI. Right: 3D display on a headset-free lenticular 3D screen. 
The video demonstration is available at: https://www.youtube.com/shorts/hJDVb2TzBr0.
}
\label{fig:9}       
\end{figure}

A visual 3D scene representation of a learning environment serves as a medium
through which the integration of STEM learning may be effectively transformed
into a tangible and meaningful experience. It has tremendous potential to
revolutionize education, yet it remains to be further explored. The free library
``altiro3D", which enables the representation of reality from a single 2D RGB image
or flat video, has also been developed in our Scientific FabLab \cite{bib11}. This software
allows the synthesis of multiple N-viewpoints images or video to create a realistic
(real-time) 3D immersive experience when displayed on a lenticular, free-view
LCD screen as shown in Figure \ref{fig:9}. altiro3D does not require heavy computing runtime
and can support a wide range of application scenarios in education and science.
The synthesis of multiple views from a single image is achieved through the
integration of machine learning together with deep learning algorithms, as well as
simple inpainting techniques, which map all image pixels.

\section{Some STEM prototypes for schools}
\label{sec:4}

To better distinguish our few exemples of hands-on STEM activities conducted in primary and secondary
Italian schools, we described next some of our STEM prototypes for schools, giving all the useful 
information available.

\subsection{Large microcontroller board replica}
\label{subsec:4a}

Microcontrollers, such as the popular Arduino Uno, can be employed in a
multitude of educational projects including robotics, the monitoring of
thermodynamic variables via sensors (e.g., humidity, temperature, pressure, etc),
RGB LED-based controllers, and GPS trackers, among others. This technology is
becoming increasingly accessible, offering a high degree of affordability and
versatility. The introduction of microcontrollers to primary school students
can be facilitated through the use of block-based coding.

A comprehensive examination of the use of microcontrollers in primary and
secondary schools has revealed that the primary obstacle to their implementation
is the limited scale of electronic components and the associated difficulty in
handling them. This has prompted us to develop a larger version of the most
commonly used microcontroller, Arduino Uno, together with a prototype board
and some electronic components with magnetic connections. The larger (9:1)
scale facilitates the handling of the attached components while maintaining the
overall set's ease of transportation. The scale model was constructed using a
laser cutter and a combination of 3D printed components, connectors and pins.

\begin{figure}[ht]
    \centering
        \includegraphics[width=0.18\textwidth]{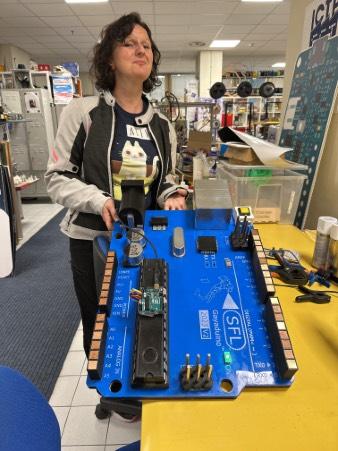}
        \includegraphics[width=0.322\textwidth]{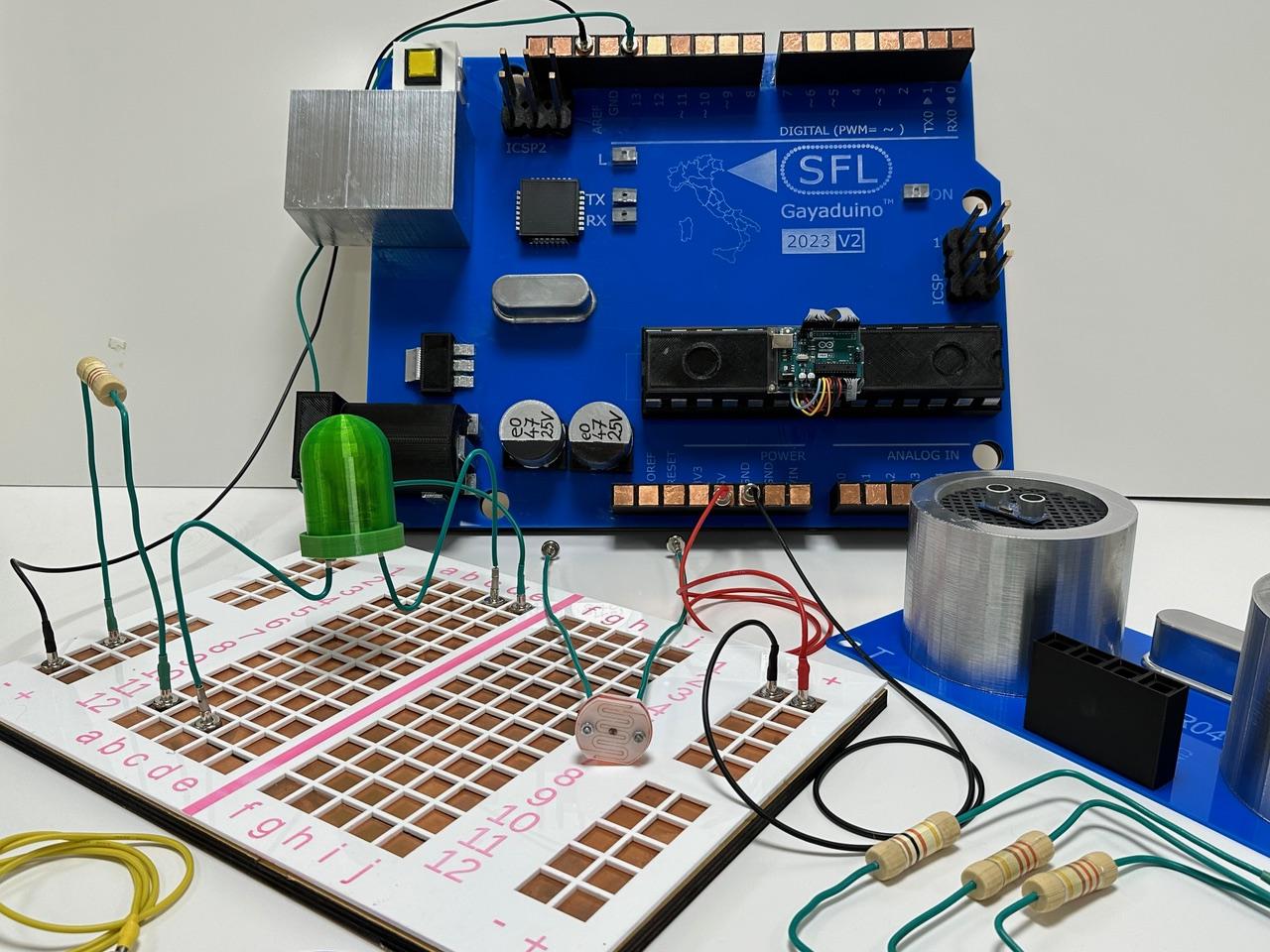}
\caption{Large microcontroller Arduino board replica.}
\label{fig:2}
\end{figure}

The fully functioning giant replica in Figure \ref{fig:2} not only allows an instructor to
explain and demonstrate how to use the microcontroller to students in a classroom
setting, but it also benefits students with visual impairments or motor skill
difficulties, enabling them to participate actively in the lectures.

\subsection{DIY lift generator for STEM: \emph{Tubora}}
\label{subsec:4b}

\begin{figure}[ht]
    \centering
        \includegraphics[width=0.2\textwidth]{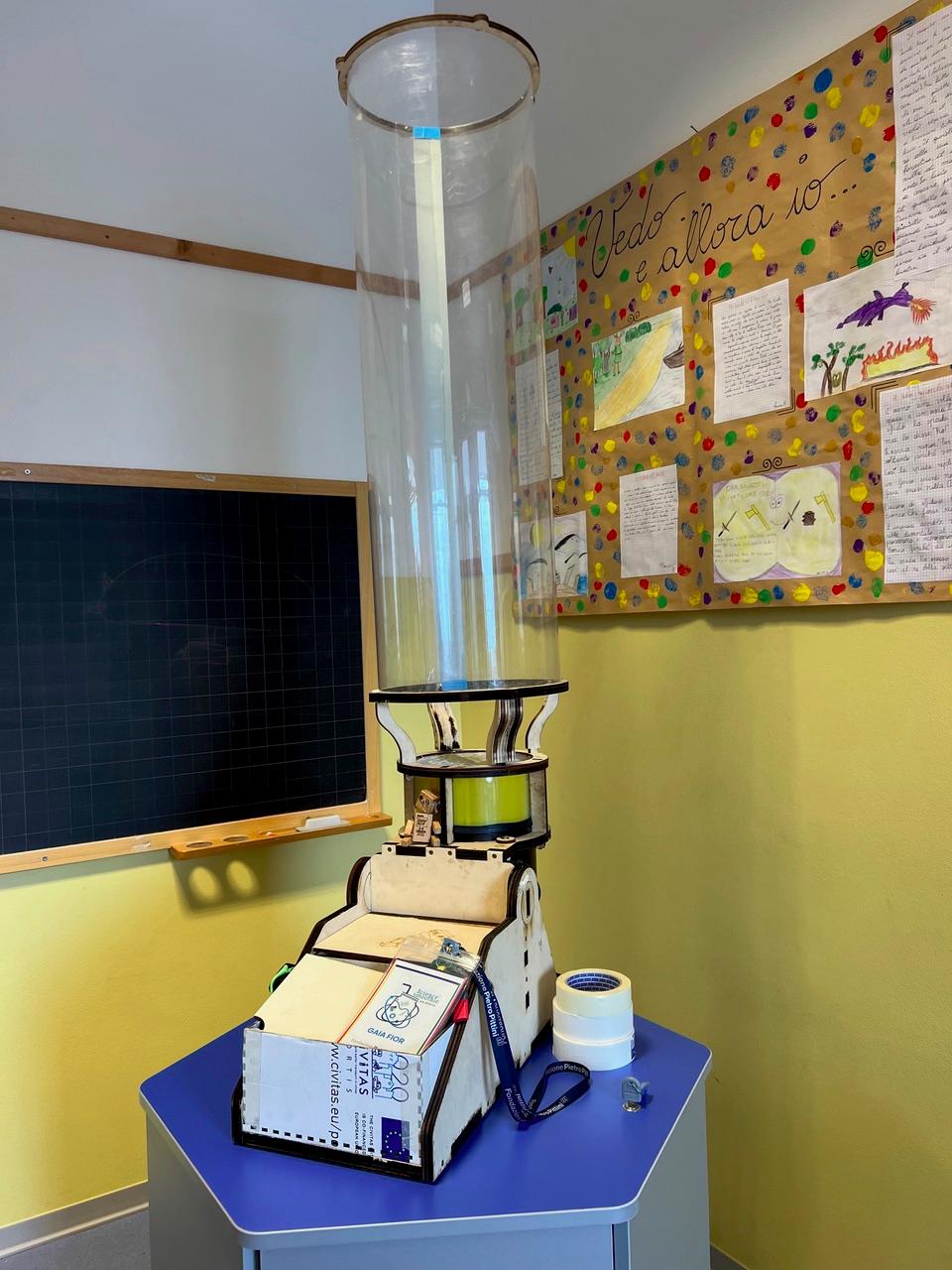}
        \includegraphics[width=0.179\textwidth]{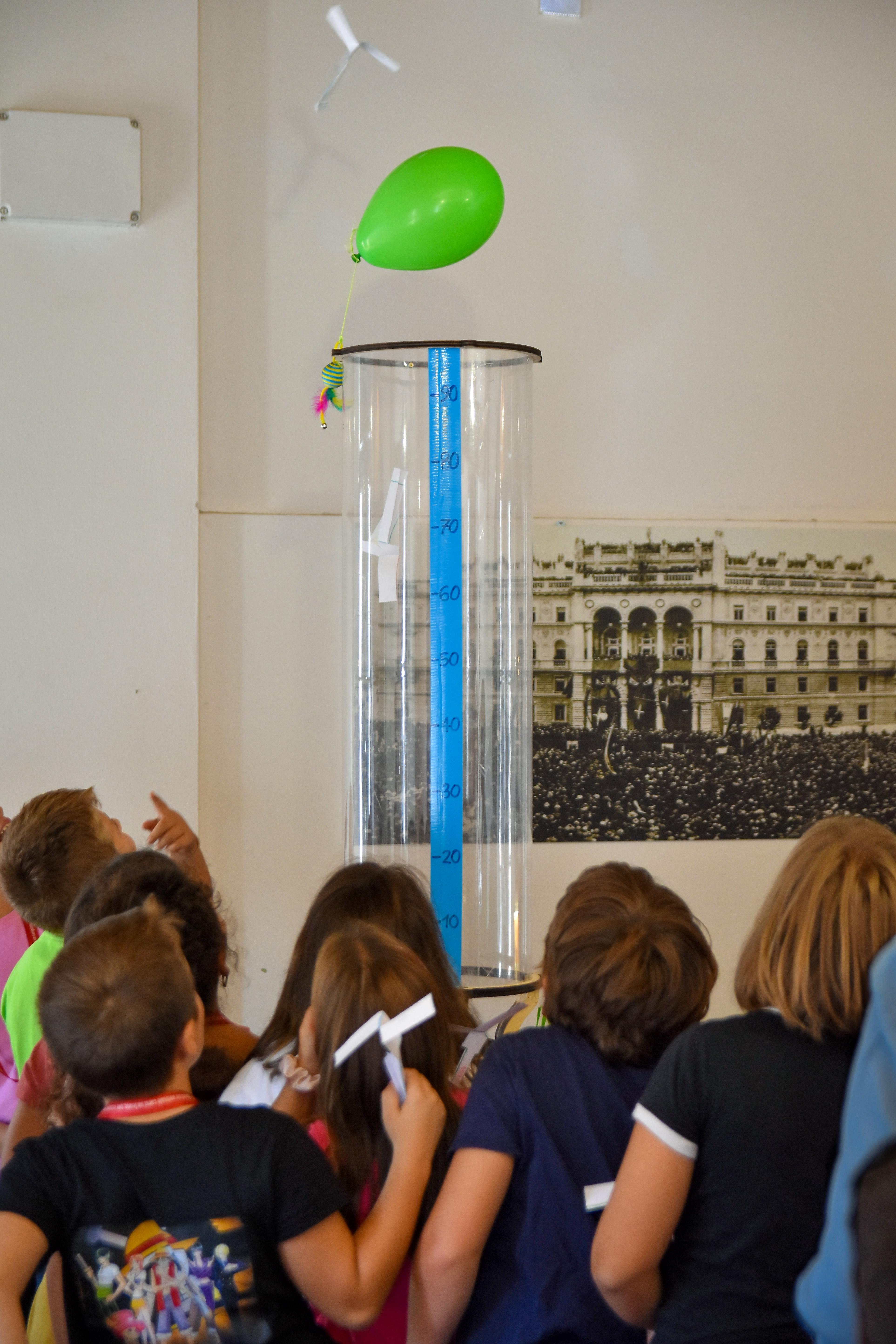}
\caption{\emph{Tubora}: An invaluable resource for students engaged in STEM studies
of aerodynamics or the mechanics of parachutes.}
\label{fig:3}
\end{figure}

In order to teach wind dynamics (testing which object flies the best and which
does not), a DIY wind tube can be built using just simple elements
such as a fan and a transparent plastic tube as shown in Figure \ref{fig:3}. This system is
designed to regulate the pull of air throughout the tube and, in addition, the tube
can be configured to point at different angles.

Placing the fan on the underside of the large tube, and inserting objects of
varying materials, students can observe the different aerodynamic characteristics
of each object. As they collect and analyze data to ascertain the mechanisms
underlying the movement of objects in a forward or downward direction, or their
retention in the air at varying angles of flow, they gain insight into the most
intriguing aspects of natural physics, namely the phenomena associated with
objects in an airflow, aerodynamics and variable drag forces.

The versatility of this exhibit is noteworthy. It allows children in the first
years of primary school (aged 6--8) to construct paper objects in a creative manner,
thereby determining the probability of them flying out of the tube. This provides
students with the opportunity to experiment with different shapes, testing them
each time, and to color them in an imaginative way. In addition to cones and
paper airplanes, students can investigate the efficacy of twirling paper
helicopters. This involves varying factors such as the type of paper, the addition
of paperclips, the size of the wings, and so forth.

\begin{figure}[ht]
    \centering
        \includegraphics[width=0.18\textwidth]{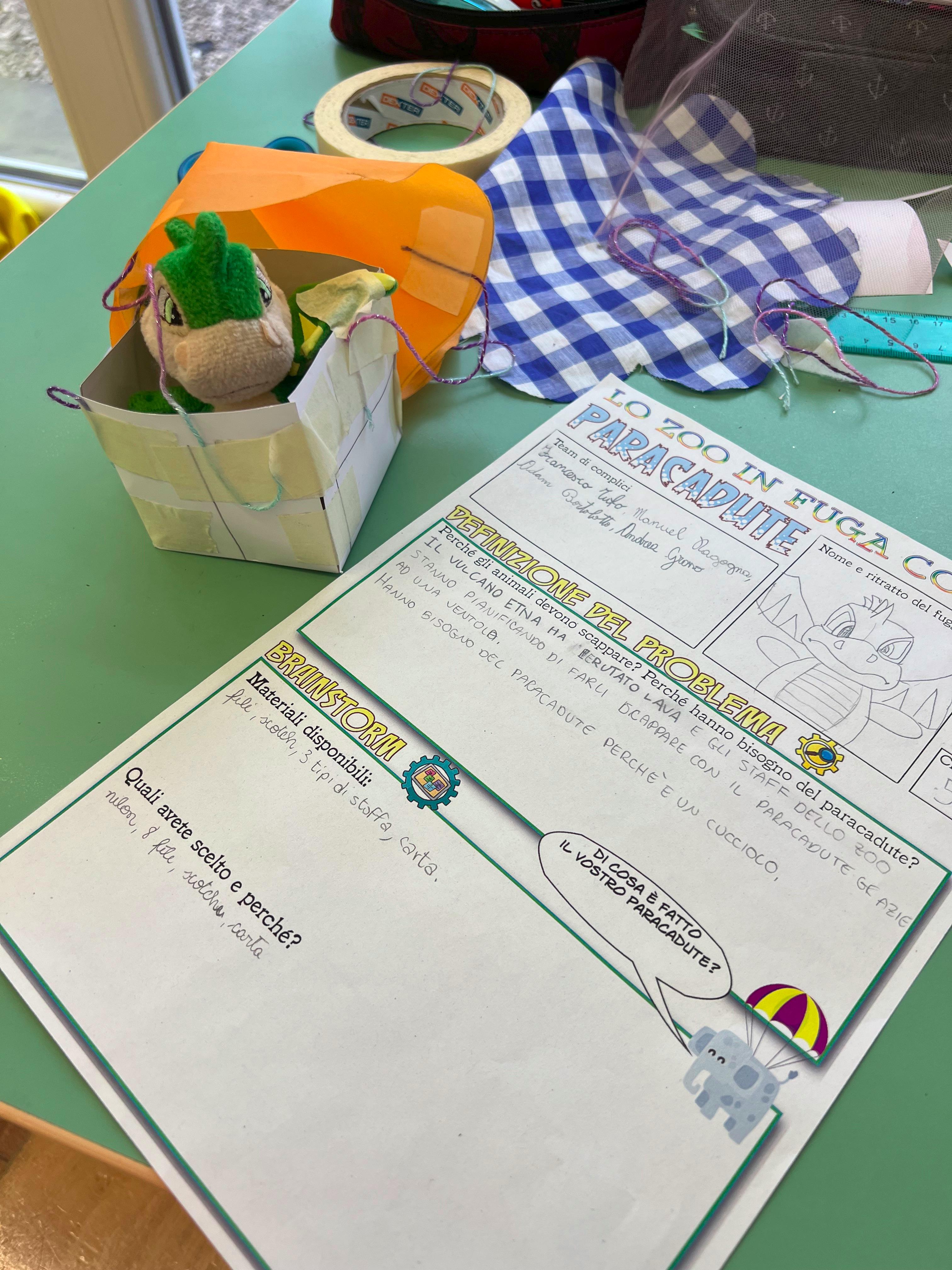}
        \includegraphics[width=0.18\textwidth]{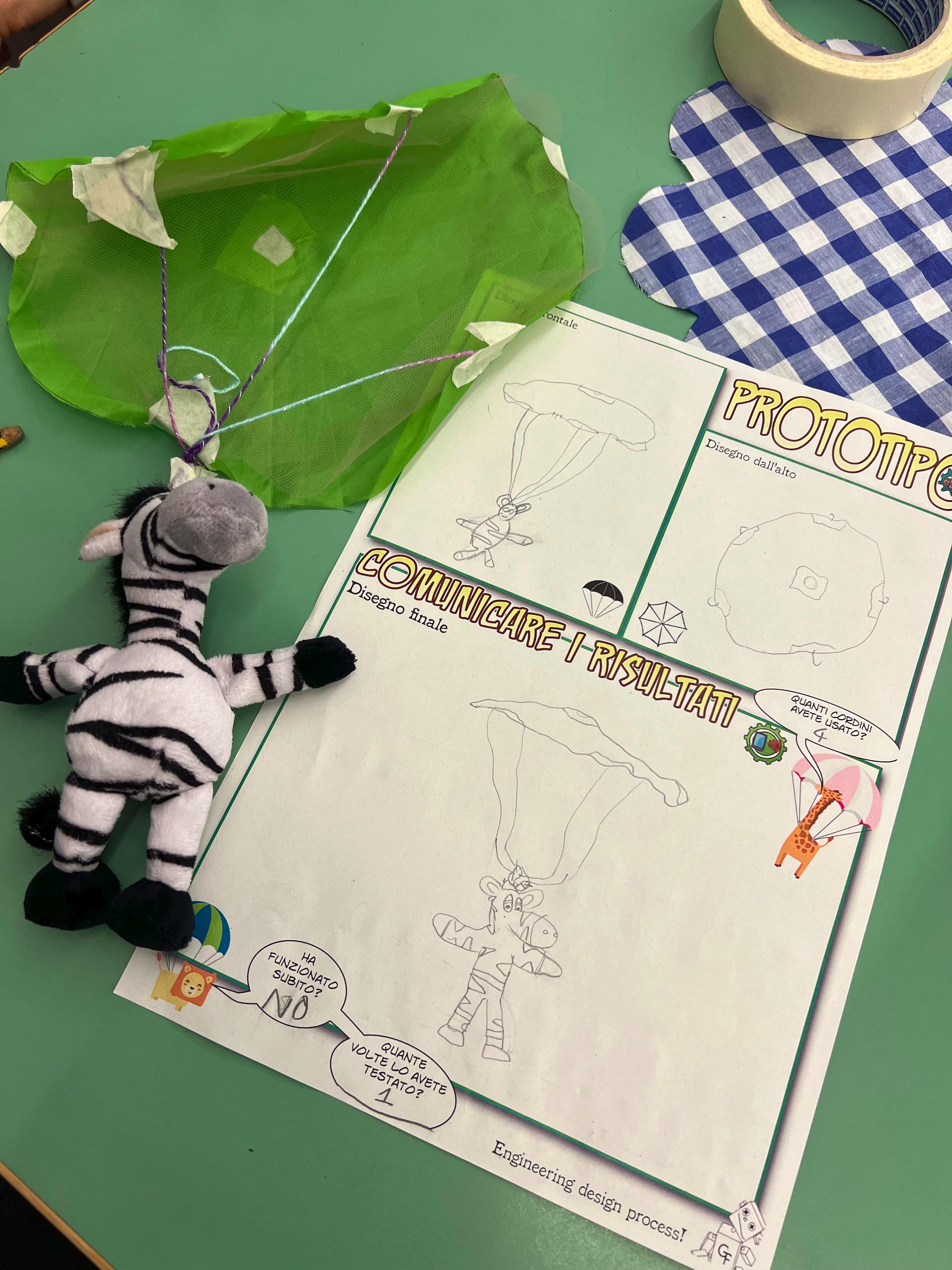}
\caption{Examples of activity sheets for the STEM Engineering activity. The Runaway Zoo.}
\label{fig:4}
\end{figure}

They are presented with an object (specifically a small, lightweight stuffed animal)
that is too heavy to be lifted by the fan alone. Each group is provided with the
same materials (paper, tulle, cotton and umbrella fabric, plastic bags, string,
masking tape) and must design the most efficient solution to allow the “skydiver”
to rise and escape from the tube. The students will have the opportunity to
experiment with different approaches to the design of the parachute canopy. They
will select the material they wish to use for the main canopy and then proceed
through the engineering design process, which involves defining the problem,
brainstorming potential solutions, creating prototypes and testing them in a wind
tunnel. Each group will also be required to complete a record sheet, which will
help them document their choices, the reasons behind them, and some
background information on the plush animal and its need to escape. They are
also required to draft sketches of the front and top views of their prototype to
complete their task thus fulfilling the final step of the engineering design process
(document and share), see Figure \ref{fig:4}. This parachute challenge offers the opportunity
to investigate a multitude of aspects related to flight and the effects of force on
an object. It also provides an opportunity to examine the engineering design
process, which is a systematic approach to problem-solving.

\section{Discussion}
\label{sec:5}

One of the objectives of this work is to include an account of
some of our experiences and activities carried out in Italian schools
using digital technologies from FabLabs.
Our school experiences and public activities 
are not only directed to school-aged childrens, but
are also being developed for interested scholars of all ages.

The current configuration of public primary and secondary schools in Italy
presents certain challenges. The minimum number of pupils per class in primary
schools is 15, with a maximum of 26 pupils permitted, except in the case of
disabled pupils. In secondary schools, the minimum number of students per class
is 18, with a maximum of 27 students permitted \cite{bib12}.

A further challenge is posed by the increasing number of students from other
countries, in the 2021/2022 academic year, 10.6\% of students in Italy were from
other countries, with the figure in the FVG region reaching 15.4\% and 22.7\%
in the province of Gorizia \cite{bib13}. A considerable number of educational institutions
are currently undergoing renewal work, with classes being temporarily relocated
to smaller classrooms or to alternative buildings that have been converted into
schools. The challenges of working in such circumstances are considerable,
particularly when the classes are overcrowded and a significant proportion of
students lack proficiency in the local language. The ages of participants in our activities
during the academic year 2023/24 were as follows. As far as laboratory hand-ons classes were 
concerned, there were 172 workshops carried out for the primary school years 1--5 
(students having 6 to 11 years of age in correspondence). The age groups consisted of 12\% 
in the 1st class, 15\% in the 2nd, 19\% in the 3rd, 24\% in the 4th, 24\% in the 5th year, 
and 6\% in multiple age groups classes.

STEM curricula are highly engaging for students, particularly those with language 
difficulties. The hands-on nature of many STEM subjects enables all students to 
participate fully, regardless of their linguistic abilities. This phenomenon is 
further evidenced when students with disabilities are present in the classroom. 
According to our experiences, STEM activities can be modified to
accommodate the specific needs of these students, with the objective of overcoming 
physical limitations and addressing mental impairments. During a lesson in which a 
class with a blind student was being taught, the possibility of combining a buzzer 
with the LED outcome of the experiment proved invaluable in fully involving the 
student in the lesson.

On another occasion, a student in a wheelchair was able to participate completely
in the activity by having all the educational materials relocated to the correct
height for him to reach. In another instance, a student with an autism spectrum
disorder had the time needed to fully appreciate the effects of air thrust on a
small parachute, while his fellow students were engaged in the collection of data
and the preparation of a subsequent demonstration.

While curricula teachers interact with students throughout the academic year,
structuring their lessons over a long time span, STEM laboratories are often
conducted by external educators with the aid of specific materials and teaching
aids, and must accommodate the needs of the school timetable. The principal
objective of these laboratories is to stimulate interest among students and
provide the curricula teachers with insights into alternative pedagogical
approaches to technical subjects. The lessons provided can then be further
developed during the school year.

The experiences of one of the authors (G.F., also an assistant in the SciFabLab) on providing 
one-off, hands-on learning activities in primary and secondary schools over the past decade 
has been predominantly focused on the delivery of engaging, 1.5 to 3-hour-long
sessions that engage the entire class, while also providing all the necessary
materials. This approach has the dual benefit of relieving teachers of the burden
of providing materials while also allowing for more focused and effective learning.

\begin{figure}[ht]
    \centering
        \includegraphics[width=0.21\textwidth]{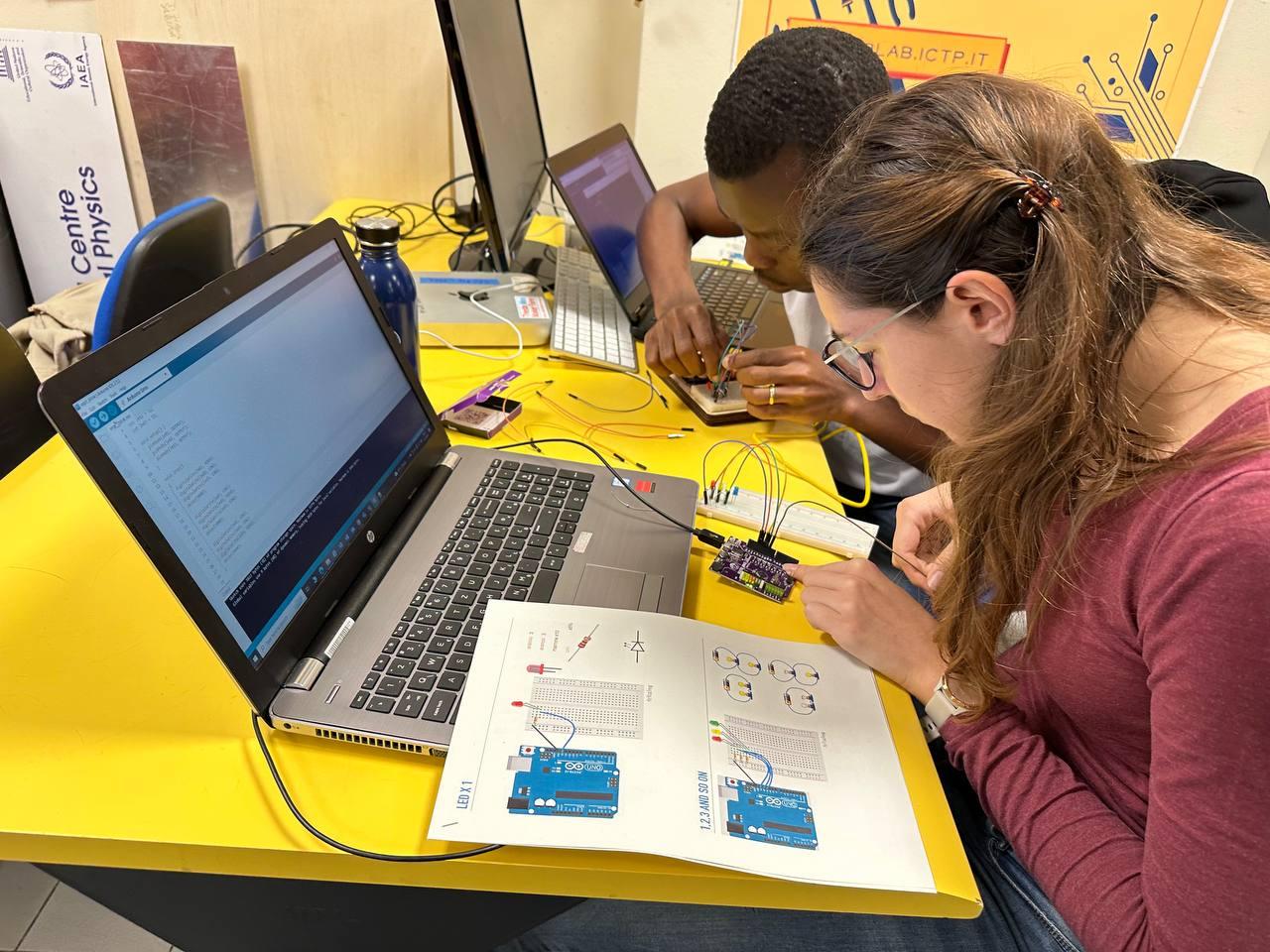}
        \includegraphics[width=0.21\textwidth]{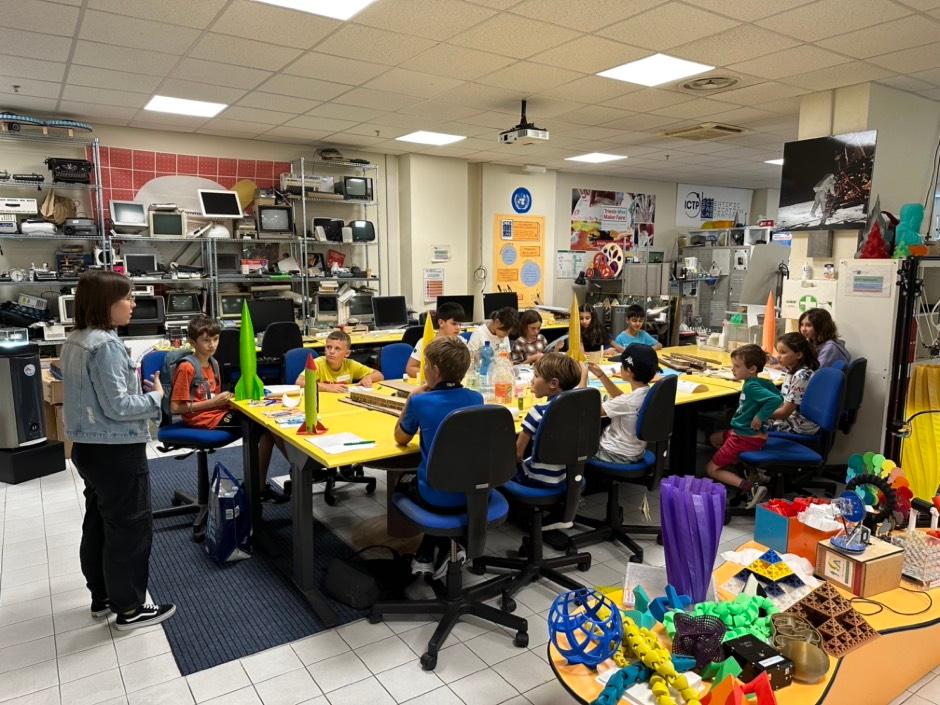}   
\caption{Students' STEM activities for Italian schools from the FVG region carried out at the Scientific 
FabLab of ICTP.}
\label{fig:10}       
\end{figure}

Students currently engaged in primary and secondary education have
experienced the challenges of the Covid-19 lockdown and a prolonged period of
distance learning, which has resulted in significant difficulties in working in
groups and reduced dexterity. This kind of learning, utilizing
hands-on experience using digital technologies, can be extremely useful in 
addressing both of these problems. In our experience, dividing the class
into four groups is optimal, as it allows for more focused and effective learning
while allowing a single educator (which is the typical situation) to
monitor the progress of each group, ensuring that all groups remain on track
while providing additional challenges to the more advanced group.

While working with smaller groups or a one-to-one approach would also be
interesting, the limitations posed by the need to engage the whole class with
a single educator limit the possibilities. An ideal configuration would have 
four groups of students, forcing them to work as teams to exchange ideas and
compare their different approaches, allowing for a more focused and effective
learning environment. Furthermore, this configuration has enabled teachers to examine the
interactions between their students in a group setting, offering valuable insights.
The decision regarding the composition of groups is always a challenging one,
with students frequently attempting to maintain proximity to their acquaintances,
thereby creating groups that are not optimally balanced. When working with
primary school pupils, the involvement of teachers, who possess a comprehensive
understanding of the students and are capable of dividing them (maintaining the
separation of more challenging students and creating more balanced groups), can
be invaluable in this regard. In the context of working with secondary school
students, it is possible to allow them to choose their own groups, thereby giving
them the opportunity to assume this responsibility.

In most cases, the curricula teachers participate, assisting students that are
having difficulty and providing important context with references to lessons already
given on similar topics. Frequently, they choose to discuss in advance the topics
covered during such activities, stimulating students' curiosity before the
laboratory, creating an optimal learning environment.
The class is often requested to provide feedback on the STEM topics presented,
which may take the form of a simple drawing or collaborative poster for primary
school students, or a summary or more creative medium such as a video, comic,
rap song, newscast, or other form for secondary students.
The variety of media employed and the creativity displayed by classes that have
previously encountered disciplinary problems during the laboratory is often
noteworthy.

A recent extensive literature review indicates that digital technology can positively affect 
STEM education in terms of knowledge or skill acquisition and learning engagement in young 
children \cite{Hu24}. This fact was found regardless of gender but highly relevant to age. Learning from, 
with, and through technology occurs when the technology is easy to be used as a tool to deliver knowledge. 
In most cases, these technologies become limited within the context of early childhood education 
since they include computer programming, simulation software, robotics, 3D technology, video games, 
apps., etc. The new trends of emerging digital technologies, such as AI, in the STEM learning context 
become limited in children's education \cite{Yan24}. An important problem outlined in these
existing approaches is the age of students. Our study differentiates with such reports because we actually 
motivate the primary and secondary schools to participate in STEM festivals. Ours are preliminary outcomes 
based on our experiences not only carried out in Italian schools but also during large, free public activities 
like Maker Faires and Science Picnics which allow anyone, of any age, to explore scientific concepts in 
simple, innovative ways. By this, we encourage the young science scholars to be creative and incentivize them
to develop their own abilities for research, discovery and ingenuity. 

Our goal in this article is to introduce readers on the importance of adopting
digitally controlled technologies found in FabLabs for tangible and visual STEM learning.
Educators may adjust their standard practices to incorporate hands-on STEM
learning activities into their curriculum at their own pace and own needs.
Not all scientific prototypes illustrated are for school-aged children, as 
with the Cloud Chamber and the AI-based 3D imaging.
We aim at getting better understanding of science through the utilization of tangible and 
visual examples as those illustrated and explained in our paper. In our study we
keep in mind the relevant work carried out
by teachers as well as aim to reach a larger range of professionals,
decision-makers, politicians, scientists and interested students in general. This
works aims to be of benefit for the progress of society as a whole.

\section{Conclusions}
\label{sec:6}

STEM students can greatly benefit when receiving training and education based
on tangible and visual learning. The prototypes presented in this work as
examples allows us to put concrete experiences in their hands. The aforementioned
examples have been generated utilizing digital technologies that are readily
accessible in any FabLab worldwide. We have discussed and assessed some of
our experiences and activities done in Italian schools and large public events 
using and demonstrating such devices such as  
touching complex 3D structures, visualizing cosmic rays, and 
building the needed blocks to learn microcontrollers in class.

At present, it is difficult to compare the similarities and differences
between our research on tangible and visual STEM learning experiences and
existing studies since there are not many similar studies.
It is premature to derive definitive conclusions because more data within different 
academic environments and experiences around the world are needed. 

We have built upon and verified how visual and tangible STEM learning emphasizes student 
potential and opens a smooth connection between 
the understanding of abstract scientific phenomena and a deeper, more
comprehensive learning experience. This novel approach in action is appealing to scholars
and can sustain new generations of young science scholars in an inclusive way facilitating 
the quality of STEM education for all.

\section*{Author contributions}

Gaia Fior: Conceptualization, Methodology creation, Data Curation, Validation, Writing - review \& editing.
Carlo Fonda: Conceptualization, Methodology creation, Investigation, Resources, Writing - review \& editing.
Enrique Canessa: Conceptualization, Methodology creation, Investigation, Formal analysis, Writing - review \& editing. Supervision.
All authors have approved the final version of the manuscript for publication.

\section*{Use of Generative-AI tools declaration}
The authors declare they have not used Artificial Intelligence (AI) tools in the creation of this article.

\subsection*{Acknowledgments}
The authors would like to thank Sara Sossi, Erika Ronchin, and Marco Baruzzo who have helped with
their ICTP SciFabLab STEM activities during the last few years. Sincere gratitude is
also due to Fondazione Pietro Pittini (FPP) for partnering with us in many of these
projects.

\section*{Conflict of interest}
The authors declare no conflict of interest in this paper.

\section*{Author's biography}

\noindent Gaia Fior (Masters in Natural Sciences) is assistant of the ICTP SciFabLab since its foundation and
organizes, and collaborates with, STEM educational projects. \\

\noindent Carlo Fonda works at the Science, Technology and Innovation (STI) Unit of the ICTP in Trieste, Italy. He is the
co-founder and manager of the ICTP SciFabLab and co-organizer of the Maker Faire Trieste since 2014. \\

\noindent Enrique Canessa (Ph.D in Physics) is with the Science, Technology and Innovation (STI) Unit of the ICTP in Trieste, Italy. He is the 
co-founder of the ICTP SciFabLab and co-organizer of the Maker Faire Trieste since 2014.

\end{document}